# Electron Landau damping in a laboratory dipole magnetic field plasma

**N.I. Grishanov[1], A.F.D. Loula[1], C.A. de Azevedo[2], J. Pereira Neto[2]**

[1]Laboratório Nacional de Computação Científica, Petrópolis, RJ, BRASIL
[2]Universidade do Estado do Rio de Janeiro, Rio de Janeiro, RJ, BRASIL

Levitated dipole configurations provide a promising approach [1, 2] for the magnetic confinement of plasmas for fusion research and planetary plasmaspheric simulations. The point dipole magnetic field is a good approximation for magnetospheric plasmas of the Earth and other planets, whereas the laboratory dipole magnetic field (with the finite current ring radius) should be considered for LDX (Levitated Dipole eXperiment) plasma [2]. As is well known, a two-dimensional kinetic wave theory in such plasmas should be based on the solution of Maxwell's equations using the correct "kinetic" dielectric tensor. In this paper, we estimate the collisionless wave dissipation of radio-frequency waves by the trapped and untrapped (circulating or passing) in a laboratory dipole magnetic field plasma (LDMFP) using the transverse and longitudinal dielectric permittivity elements evaluated in Ref. [3].

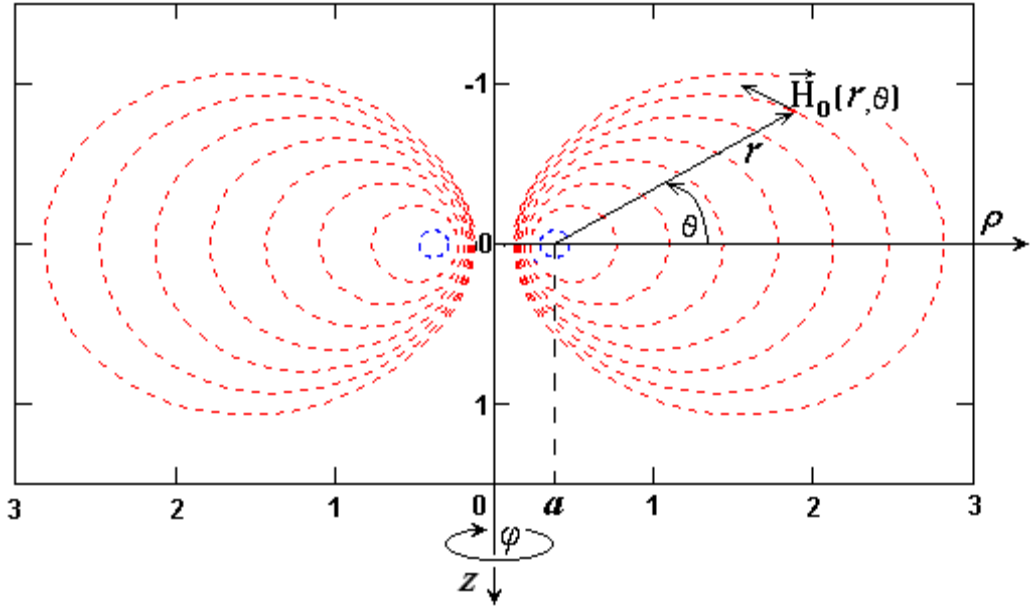

*Fig. 1. The cylindrical ($\rho, \phi, z$) and quasi-toroidal ($r, \theta, \phi$) coordinates for a laboratory dipole configuration.*

To describe a 2D axisymmetric LDMFP, see Fig. 1, we use the quasi-toroidal coordinates ($r, \theta, \phi$) connected with cylindrical ones ($\rho, \phi, z$) as

$$\rho = a + r\cos\theta, \qquad z = -r\sin\theta, \qquad \phi = \phi.$$

In this case, the cylindrical components of an equilibrium magnetic field, $\mathbf{H}_0$, are

$$H_\rho = \frac{2Ir\sin\theta}{c(a+r\cos\theta)\sqrt{r^2+4a^2+4ar\cos\theta}}\left[K(\kappa) - \frac{r^2+2a^2+2ar\cos\theta}{r^2}E(\kappa)\right] \qquad (1)$$

$$H_\phi = 0, \qquad H_z = \frac{2I}{c\sqrt{r^2+4a^2+4ar\cos\theta}}\left[K(\kappa) - \frac{r+2a\cos\theta}{r}E(\kappa)\right] \qquad (2)$$

where $a$ is the ring radius, $I$ is the ring current, $c$ is the speed of light,

$$K(\kappa) = \int_0^{\frac{\pi}{2}} \frac{dx}{\sqrt{1-\kappa \sin^2 x}}, \qquad E(\kappa) = \int_0^{\frac{\pi}{2}} \sqrt{1-\kappa \sin^2 x}\, dx, \qquad \kappa = \frac{4a(a+r\cos\theta)}{r^2 + 4a^2 + 4ar\cos\theta}. \qquad (3)$$

Introducing the new variables $(v, \mu, L)$ instead of $(v_\parallel, v_\perp, r)$ as

$$v = \sqrt{v_\parallel^2 + v_\perp^2}, \qquad \mu = \frac{v_\perp^2 H_0(r,0)}{v^2 H_0(r,\theta)}, \qquad L = \frac{\pi a}{\sqrt{r^2 + 4a^2 + 4ar\cos\theta}\,[(2-\kappa)K(\kappa) - 2E(\kappa)]} \qquad (4)$$

where $H_0 = \sqrt{H_\rho^2 + H_z^2}$ is the module of $\mathbf{H}_0$, the Vlasov equation for the first harmonics of the perturbed distribution function, $f(t,\mathbf{r},\mathbf{v}) = \sum_s^{\pm 1} \sum_l^{\pm\infty} f_l^s(L,\theta,v,\mu)\exp(-i\omega t + in\phi + il\sigma)$, in the zeroth order of a magnetization parameter, can be reduced to the first order differential equation with respect to one $\theta$ variable

$$\sqrt{1 - \frac{\mu}{b(L,\theta)}}\, \frac{1}{\lambda(L,\theta)}\, \frac{\partial f_l^s}{\partial \theta} - is\frac{La}{v}(\omega + l\Omega_c) f_l^s = Q_l^s. \qquad (5)$$

Here 
$$\lambda(L,\theta) = \frac{cr^2 H_0(r,\theta)(a+r\cos\theta)\sqrt{r^2 + 4a^2 + 4ar\cos\theta}}{2ILa[(r^2 + 2a^2 + 3ar\cos\theta)E(\kappa) - r(r+a\cos\theta)K(\kappa)]} \qquad (6)$$

$$b(L,\theta) = \frac{H_0(r(L,0),0)}{H_0(r(L,\theta),\theta)}, \quad \Omega_c = \frac{eH_0}{Mc}, \quad F = \frac{N}{\pi^{1.5} v_T^3}\exp\left(-\frac{v^2}{v_T^2}\right), \quad v_T = \sqrt{\frac{2T}{M}} \qquad (7)$$

$$Q_0^s = \frac{e}{T} LaF \sqrt{1 - \frac{\mu}{b(L,\theta)}}\, E_\parallel, \qquad Q_{\pm 1}^s = \frac{se}{2T} LaF \sqrt{\frac{\mu}{b(L,\theta)}}(E_n \mp iE_b) \qquad (8)$$

where $l = 0,\pm 1$ is the number of the cyclotron harmonics, $\sigma$ is the gyrophase angle in velocity space; $F$ is the equilibrium distribution function of particles with the density $N$, mass $M$, charge $e$, temperature $T$; $E_n$, $E_b$ and $E_\parallel$ are, respectively, the normal, binormal and parallel perturbed electric field components relative to $\mathbf{H}_0$; further $E_{\pm 1} = E_n \mp iE_b$. The index of particles kind is omitted in Eqs. (5,7,8). By $s = \pm 1$ we differ the particles with positive and negative parallel velocity, $v_\parallel$, relative to $\mathbf{H}_0$: $v_\parallel = sv\sqrt{1 - \mu/b(L,\theta)}$. Note, the "old" $r$ variable in Eqs. (6) and (7) should be determined by $r(L,\theta)$ satisfying Eq. (4), $L=L(r,\theta)$.

Since LDMFP is a configuration with one minimum of $\mathbf{H}_0$, the plasma particles should be split in the two populations of the so-called trapped and untrapped particles. In the phase volume such separation can be done by the $\mu$ variable, see Fig.2. In dependence on $\mu$ and $\theta$ we have 1) $0 \leq \mu \leq \mu_0$, $-\pi \leq \theta \leq \pi$ for the untrapped particles, where $\mu_0 = b(L,\pi)$, and 2) $\mu_0 \leq \mu \leq 1$, $-\theta_t \leq \theta \leq \theta_t$ for the trapped particles, where $\pm\theta_t(\mu,L)$ are the reflection points (or stop points, or mirror points) of the trapped particles, i.e., $v_\parallel(v,\mu,L,\pm\theta_t) = 0$.

After solving Eq. (6), the 2D transverse and longitudinal (relative to $\mathbf{H}_0$) current density components, respectively $j_{\pm 1}$ and $j_\parallel$, can be found as

$$j_\parallel(L,\theta) = \frac{\pi e}{b(L,\theta)} \sum_s^{\pm 1} s \int_0^\infty v^3 dv \left[ \int_0^{\mu_0} f_{0,u}^s(L,\theta,v,\mu)d\mu + \int_{\mu_0}^{b(L,\theta)} f_{0,t}^s(L,\theta,v,\mu)d\mu \right] \qquad (9)$$

$$j_l(L,\theta) = \frac{\pi e}{2b(L,\theta)} \sum_s^{\pm 1} \int_0^\infty v^3 dv \left[ \int_0^{\mu_0} \frac{f_{l,u}^s \sqrt{\mu}\, d\mu}{\sqrt{b(L,\theta) - \mu}} + \int_{\mu_0}^{b(L,\theta)} \frac{f_{l,t}^s \sqrt{\mu}\, d\mu}{\sqrt{b(L,\theta) - \mu}} \right], \quad l = \pm 1 \qquad (10)$$

where the subscribed indexes $u$ and $t$ correspond to the untrapped and trapped particles, respectively. Of course, the population of $u$-particles is very small at the external magnetic surfaces since $\mu_0 \to 0$ if $L$ increases. Note, the normal and binormal to $\mathbf{H}_0$ current density components, $j_n$ and $j_b$ in our notation, are equal to $j_n = j_{+1} + j_{-1}$ and $j_b = i(j_{+1} - j_{-1})$.

The expressions for $j_l|_{l=\pm 1}$ are convenient to analyze the cyclotron resonance effects on the fundamental cyclotron frequency of both the ions (if $l=-1$) and electrons (if $l=+1$) in the explicit form.

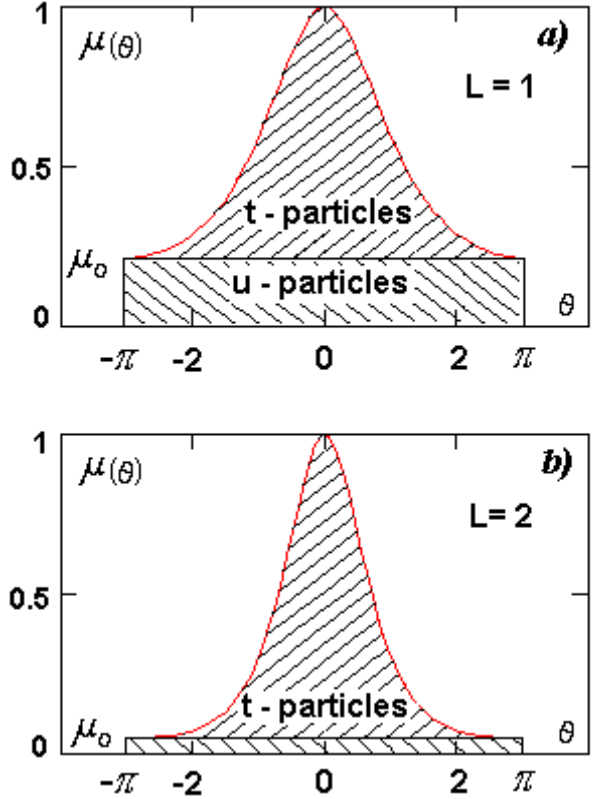

*Fig. 2. The domain of the trapped and untrapped particles in a laboratory dipole plasma model.*

To describe the bounce-periodic motion of the trapped and untrapped particles along $H_0$-field line, it is convenient to introduce the new time-like variable $\tau$ instead of $\theta$,

$$\tau(\theta) = \int_0^\theta \frac{\lambda(L,\theta')}{\sqrt{1-\mu/b(L,\theta')}} d\theta', \qquad (11)$$

accounting for that the bounce-periods of the $u$- and $t$-particles are proportional to $T_{b,u}=2\tau(\pi)$ and $T_{b,t}=4\tau(\theta_\tau)$. After this, the distribution functions of $u$- and $t$-particles can be defined by the corresponding Fourier series ($\alpha=u,t$)

$$f_{l,\alpha}^s = \sum_p^{\pm\infty} f_{l,\alpha}^{s,p} \exp\left[ ip\frac{2\pi}{T_{b,\alpha}}\tau(\theta) + isl\frac{La}{v}\int_0^\theta \frac{\tilde\Omega_{c,\alpha}\lambda(L,\theta')}{\sqrt{1-\mu/b(L,\theta')}} d\theta' \right] \qquad (12)$$

where $\tilde\Omega_{c,u} = \Omega_{c,u} - \overline\Omega_{c,u}$ and $\tilde\Omega_{c,t} = \Omega_{c,t} - \overline\Omega_{c,t}$ are the oscillating parts of the cyclotron frequencies of the $u$- and $t$-particles; and $\overline\Omega_{c,u}$ and $\overline\Omega_{c,t}$ are the corresponding bounce-averaged cyclotron frequencies (secular part of $\Omega_c$) of untrapped and trapped particles:

$$\overline\Omega_{c,u} = \frac{2}{T_{b,u}} \int_0^\pi \frac{\Omega_c \lambda(L,\theta)}{\sqrt{1-\mu/b(L,\theta)}} d\theta, \qquad \overline\Omega_{c,t} = \frac{4}{T_{b,t}} \int_0^{\theta_t} \frac{\Omega_c \lambda(L,\theta)}{\sqrt{1-\mu/b(L,\theta)}} d\theta . \qquad (13)$$

To evaluate the dielectric tensor elements we use the Fourier expansions of the perturbed electric field and current density components over $\theta$

$$b(L,\theta)\mathbf{j}(L,\theta) = \sum_{m}^{\pm\infty} \mathbf{j}^{(m)}(L)\exp(im\theta), \qquad \lambda(L,\theta)\mathbf{E}(L,\theta) = \sum_{m'}^{\pm\infty} \mathbf{E}^{(m')}(L)\exp(im'\theta). \quad (14)$$

As a result,

$$\frac{4\pi i}{\omega} j_l^{(m)} = \frac{2i}{\omega}\int_{-\pi}^{\pi} b(L,\theta) j_l(L,\theta)\exp(-im\theta)d\theta = \sum_{m'}^{\pm\infty}\left(\varepsilon_{l,u}^{m,m'} + \varepsilon_{l,t}^{m,m'}\right) E_l^{(m')}, \quad l=\pm 1 \quad (15)$$

$$\frac{4\pi i}{\omega} j_\parallel^{(m)} = \frac{2i}{\omega}\int_{-\pi}^{\pi} b(L,\theta) j_\parallel(L,\theta)\exp(-im\theta)d\theta = \sum_{m'}^{\pm\infty}\left(\varepsilon_{\parallel,u}^{m,m'} + \varepsilon_{\parallel,t}^{m,m'}\right) E_\parallel^{(m')} \quad (16)$$

where $\varepsilon_{l,u}^{m,m'}$, $\varepsilon_{l,t}^{m,m'}$ and $\varepsilon_{\parallel,u}^{m,m'}$, $\varepsilon_{\parallel,t}^{m,m'}$ are the contribution of untrapped ($u$) and trapped ($t$) particles to the transverse and longitudinal permittivity elements, respectively. After the $s$-summation, the dielectric permittivity elements can be expressed as

$$\varepsilon_{l,u}^{m,m'} = \frac{\omega_p^2 L\, a}{8\omega\pi^{2.5} v_T} \sum_{p=-\infty}^{\infty}\int_0^{\mu_0}\mu\, d\mu \int_{-\infty}^{\infty}\frac{u^4\exp(-u^2)}{pu-Z_{l,u}} A_{l,p}^m(u,\mu) A_{l,p}^{m'}(u,\mu)du \quad (17)$$

$$\varepsilon_{l,t}^{m,m'} = \frac{\omega_p^2 L\, a}{8\omega\pi^{2.5} v_T} \sum_{p=-\infty}^{\infty}\int_{\mu_0}^{1}\mu\, d\mu \int_{-\infty}^{\infty}\frac{u^4\exp(-u^2)}{pu-Z_{l,t}} \hat{B}_{l,p}^m(u,\mu) B_{l,p}^{m'}(u,\mu)du \quad (18)$$

$$\varepsilon_{\parallel,u}^{m,m'} = \frac{\omega_p^2 L^2 a^2}{8\pi^3 v_T^2}\sum_{p=1}^{\infty}\frac{1}{p^2}\int_0^{\mu_o} T_{b,u}\left(C_p^m C_p^{m'} + C_{-p}^m C_{-p}^{m'}\right)\left[1+2u_p^2+2i\sqrt{\pi}u_p^3 W(u_p)\right]d\mu \quad (19)$$

$$\varepsilon_{\parallel,t}^{m,m'} = \frac{\omega_p^2 L^2 a^2}{8\pi^3 v_T^2}\sum_{p=1}^{\infty}\frac{1}{p^2}\int_{\mu_0}^{1} T_{b,t} D_p^m D_p^{m'}\left[1+2v_p^2+2i\sqrt{\pi}v_p^3 W(v_p)\right]d\mu . \quad (20)$$

Here we have introduced the following definitions

$$A_{l,p}^m(u,\mu) = \int_{-\pi}^{\pi}\cos\left[m\theta - \left(\frac{2\pi p}{T_{b,u}} - \frac{lLa}{uv_T}\overline{\Omega}_{c,u}\right)\tau(\theta) - \frac{lLa}{uv_T}\int_0^{\theta}\frac{\Omega_c\lambda(L,\eta)d\eta}{\sqrt{1-\mu/b(L,\eta)}}\right]\frac{d\theta}{\sqrt{b(L,\theta)-\mu}} \quad (21)$$

$$\hat{B}_{l,p}^m(u,\mu) = \int_{-\theta_t}^{\theta_t}\cos\left[m\theta - \left(\frac{2\pi p}{T_{b,t}} - \frac{lLa}{uv_T}\overline{\Omega}_{c,t}\right)\tau(\theta) - \frac{lLa}{uv_T}\int_0^{\theta}\frac{\Omega_c\lambda(L,\eta)d\eta}{\sqrt{1-\mu/b(L,\eta)}}\right]\frac{d\theta}{\sqrt{b(L,\theta)-\mu}} \quad (22)$$

$$B_{l,p}^m(u,\mu) = \hat{B}_{l,p}^m(u,\mu) + (-1)^p \hat{B}_{l,-p}^m(-u,\mu), \qquad u = \frac{v}{v_T}, \qquad \omega_p^2 = \frac{4\pi N e^2}{M} \quad (23)$$

$$Z_{l,u} = \frac{LaT_{b,u}}{2\pi v_T}(\omega + l\overline{\Omega}_{c,u}), \quad Z_{l,t} = \frac{LaT_{b,t}}{2\pi v_T}(\omega + l\overline{\Omega}_{c,t}), \quad u_p = \frac{\omega La T_{b,u}}{2\pi p v_T}, \quad v_p = \frac{\omega La T_{b,t}}{2\pi p v_T} \quad (24)$$

$$C_p^m(\mu) = \int_{-\pi}^{\pi}\cos\left[m\theta - \frac{2\pi p}{T_{b,u}}\tau(\theta)\right]d\theta, \qquad W(z) = \exp(-z^2)\left[1+\frac{2i}{\sqrt{\pi}}\int_0^z \exp(t^2)dt\right] \quad (25)$$

$$D_p^m(\mu) = \int_{-\theta_t}^{\theta_t}\cos\left[m\theta - \frac{2\pi p}{T_{b,t}}\tau(\theta)\right]d\theta + (-1)^{p-1}\int_{-\theta_t}^{\theta_t}\cos\left[m\theta + \frac{2\pi p}{T_{b,t}}\tau(\theta)\right]d\theta . \quad (26)$$

Using the bounce-periodicity of the perturbed distribution functions of untrapped and trapped particles over the $\tau$–variable, the wave-particle resonance conditions in the LDMFP are defined as usually by the zeros of the corresponding denominators of the $f_{l,u}^s$ and $f_{l,t}^s$ harmonics of the perturbed distribution functions of the untrapped and trapped particles. These resonant denominators are presented in Eqs. (17) and (18) in the explicit form.

One of the main mechanisms of the radio-frequency plasma heating is the electron Landau damping of waves due to the Cherenkov resonance interaction of the parallel electric field component $E_\parallel$ with untrapped and trapped electrons. The Cherenkov resonance conditions are different for trapped and untrapped particles in the LDMFP and have nothing in common with the wave-particle resonance condition in the plasma confined in the straight magnetic field. Another important feature of LDMFP is the fact that, due to 2D $\mathbf{H}_0$-field nonuniformity, the whole spectrum of $E$-field is present in the given $m$-th harmonic of the current density, see Eqs. (15) and (16). As a result, the dissipated wave power by the trapped and untrapped electrons, $P_L = \text{Re}(E_\parallel) \cdot \text{Re}(j_\parallel)$, can be estimated by the expression

$$P_L = P_u + P_t = \frac{\omega}{8\pi} \sum_m^{\pm\infty} \sum_{m'}^{\pm\infty} \left( \text{Im}\,\varepsilon_{\parallel,u}^{m,m'} + \text{Im}\,\varepsilon_{\parallel,t}^{m,m'} \right) \left( \text{Re}\,E_\parallel^{(m)} \text{Re}\,E_\parallel^{(m')} + \text{Im}\,E_\parallel^{(m)} \text{Im}\,E_\parallel^{(m')} \right) \quad (27)$$

where, as follows from Eqs. (19) and (20)

$$\text{Im}\,\varepsilon_{\parallel,u}^{m,m'} = \frac{\omega_p^2 L^2 a^2}{4\pi^{2.5} v_T^2} \sum_{p=1}^{\infty} \frac{1}{p^2} \int_0^{\mu_0} T_{b,u} \left( C_p^m C_p^{m'} + C_{-p}^m C_{-p}^{m'} \right) u_p^3 \exp(-u_p^2) d\mu \quad (28)$$

$$\text{Im}\,\varepsilon_{\parallel,t}^{m,m'} = \frac{\omega_p^2 L^2 a^2}{4\pi^{2.5} v_T^2} \sum_{p=1}^{\infty} \frac{1}{p^2} \int_{\mu_0}^{1} T_{b,t} D_p^m D_p^{m'} v_p^3 \exp(-v_p^2) d\mu \quad (29)$$

are the contribution of untrapped, $\text{Im}\,\varepsilon_{\parallel,u}^{m,m'}$, and trapped, $\text{Im}\,\varepsilon_{\parallel,t}^{m,m'}$, particles to the imaginary part of the longitudinal permittivity elements: $\text{Im}\,\varepsilon_\parallel^{m,m'} = \text{Im}\,\varepsilon_{\parallel,u}^{m,m'} + \text{Im}\,\varepsilon_{\parallel,t}^{m,m'}$. Thus, we see that for the given $\omega$, $m$, $n$, $L$ and $E$-field amplitudes the parts $P_u$ and $P_t$ differ by the different contributions of untrapped and trapped electrons to $\text{Im}\,\varepsilon_\parallel^{m,m'}$.

There is another important plasma heating mechanism due to the cyclotron wave damping in the range of ion or/and electron cyclotron frequencies, when the plasma particles interact effectively with the transverse electric field components, $E_l = E_n - ilE_b$, $l = \pm 1$. Under the cyclotron resonance heating (on the fundamental harmonics) the wave power absorbed, $P_{C,l} = \text{Re}(E_l \cdot j_l^*)$, can be expressed as

$$P_{C,l} = P_{l,u} + P_{l,t} = \frac{\omega}{8\pi} \sum_m^{\pm\infty} \sum_{m'}^{\pm\infty} \left( \text{Im}\,\varepsilon_{l,u}^{m,m'} + \text{Im}\,\varepsilon_{l,t}^{m,m'} \right) \left( \text{Re}\,E_l^{(m)} \text{Re}\,E_l^{(m')} + \text{Im}\,E_l^{(m)} \text{Im}\,E_l^{(m')} \right). \quad (30)$$

As was mentioned above, $l=-1$ corresponds to the wave dissipation under the ion-cyclotron resonance heating when $\omega \approx \Omega_{c,i}$; whereas $l=1$ should be used under the electron-cyclotron plasma heating when $\omega \approx |\Omega_{c,e}|$. The contributions of untrapped and trapped particles to $\text{Im}\,\varepsilon_{l,u}^{m,m'}$ and $\text{Im}\,\varepsilon_{l,t}^{m,m'}$ can be readily derived from Eqs. (17) and (18) by using the well known residue (or Landau rule) method.

Now, let us calculate numerically the contributions of untrapped and trapped electrons to the imaginary parts of the diagonal ($m=m'$) parallel permittivity elements, $\text{Im}\,\varepsilon_{\parallel,u}^{m,m}$ and $\text{Im}\,\varepsilon_{\parallel,t}^{m,m}$ by Eqs. (28 and 29), versus $\omega$ in a laboratory dipole plasma with the main parameters: $a=0.35$ m, $T(r)=200/L^{8/3}$ eV, $N(r)=10^{19}/L^4$ m$^{-3}$, corresponding to LDX plasma parameters presented in Ref. [4]. The computations are carried out for waves with a mode-number $m=2$ in a wide frequency range at the magnetic surface $L=2$.

By comparing $\text{Im}\,\varepsilon_{\parallel,u}^{m,m}$ and $\text{Im}\,\varepsilon_{\parallel,t}^{m,m}$ in Fig. 3, we see that the low frequency waves dissipate mainly due to their bounce resonant interaction with the trapped electrons. The maximum of the wave dissipation (by $\text{Im}\,\varepsilon_{\parallel,t}^{m,m}$) is reached under the conditions when the wave frequency is comparable with the bounce-frequency of the trapped electrons, $\omega \sim \omega_{b,t}$. In this case, the low numbers ($p=1$, $p=2$) of the bounce resonances are principal and give the main contributions to $\text{Im}\,\varepsilon_\parallel^{m,m'}$. Moreover, the bounce resonances with low $p$ should be principal considering the waves with low mode-numbers ($m=2$ in our case), as was shown in a tokamak plasma [16-18], the more effective wave-particle interactions take place when the bounce resonance numbers and poloidal mode-numbers are comparable, $p \sim m$.

As shown in Fig. 3, wave dissipation by the untrapped particles/electrons in LDMFP is always small because of their small population in the phase volume, and should be smaller at the external magnetic surfaces since $\mu_0 \to 0$ if $L$ increases. Nonetheless, the maximum of $\text{Im}\,\varepsilon_{\parallel,u}^{m,m}$ is reached (by analogy with the trapped particles) under the conditions when $\omega \sim \omega_{b,u}$ and the phase wave velocity, $v_{ph} = \omega L a / m$, is of the order of the electron thermal velocity, $v_{ph} \sim v_T$.

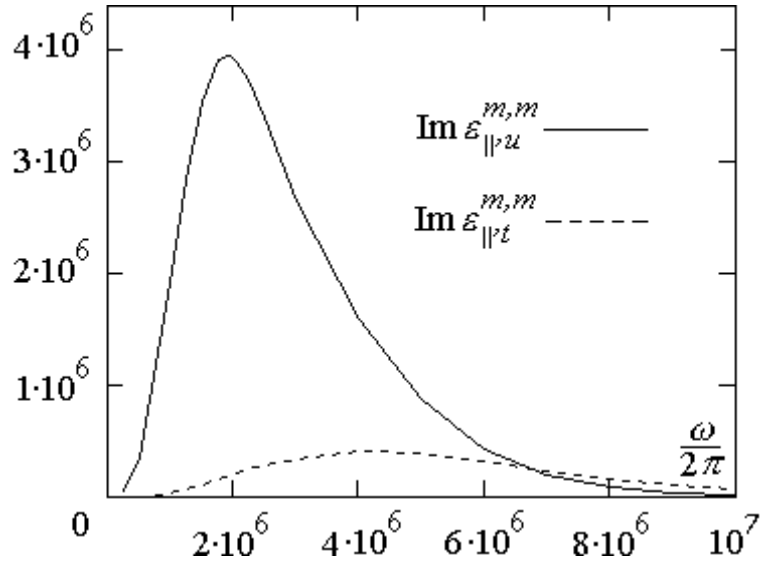

*Fig. 3. The contributions of the trapped ('solid' curve) and untrapped ('dashed' curve) electrons to the imaginary parts of the parallel permittivity elements in dependence of the wave frequency.*

Note that, $\text{Im}\,\varepsilon_{\parallel,u}^{m,m}$ and $\text{Im}\,\varepsilon_{\parallel,t}^{m,m}$ are very (exponentially) small for waves with $\omega \gg \omega_{b,t}$ and $v_{ph} \gg v_T$. However, the wave dissipation by the untrapped electrons can be larger (in a few times) than it is by the trapped particles in the frequency range where $\omega / \omega_{b,t} \sim 3 \div 4$.

In conclusion, note that analyzing the wave dissipation in LDMFP, one should remember about the other plasma heating mechanisms such as the Transit Time Magnetic Pumping (TTMP) and interaction of $E_\parallel$ and $E_t$ (by the cross-off terms of the dielectric tensor). To describe these effects in our 2D plasma model the all nine dielectric tensor components should be derived accounting for the finite *beta* and finite Larmor radius corrections. However, this is a topic of additional investigations.

*Acknowledgements:* This research was supported by CNPq of Brazil (Conselho Nacional de Desenvolvimento Científico e Tecnológico, project PCI-LNCC/MCT 382042/04-2).

**References**
[1] A. Hasegawa, *Comments on Plasma Phys. Controlled Fusion*, **1**, 147 (1987).
[2] J. Kesner, L. Bromberg, M.E. Mauel, D.T. Garnier, *17th Int. Conf. Plasma Phys. Control. Fusion Res.,* Yokahama, Japan, 19-23 October 1998, paper IAEA-F1-CN-69/ICP/09.
[3] N.I. Grishanov, A.F.D. Loula, C.A. de Azevedo, J.P. Neto, *AIP Conference Proceedings 669, 11th International Congress on Plasma Physics: ICPP2002,* p. 467, June 2003.
[4] Garnier D.T., Mauel M., Kesner J. *et al, 41-st Annual meeting of the Division of Plasma Physics* of the American Physical Society, Seattle, Washington, November 15, 1999.